\documentclass[twoside]{article}
\usepackage{Proc_NTSE_16}

\usepackage{bm}
\begin{document}
\thispagestyle{plain}
\publref{khokhlov}

\begin{center}
{\Large \bf \strut
The deuteron wave function and neutron form factors from\\
the elastic electron deuteron  scattering
\strut}\\
\vspace{10mm}
{\large \bf
Khokhlov, N. A.}
\end{center}

\noindent{
\small \it Komsomolsk-na-Amure State Technical University}

\markboth{
Khokhlov, N. A.}
{
The deuteron wave function and neutron form factors}

\begin{abstract}
A study of the deuteron structure in the framework of relativistic quantum mechanics is presented. Point-form (PF) relativistic quantum mechanics (RQM) is
applied to the elastic $eD$ scattering. The  deuteron wave function and neutron form factors are fitted to the electromagnetic deuteron form factors. We also compare results obtained by different realistic
deuteron wave functions stemming from Nijmegen-I, Nijmegen-II, JISP16, CD-Bonn, Paris, Argonne, Idaho and
Moscow (with forbidden states) potentials. It is shown that the electromagnetic deuteron form factors may be described without exchange currents.
\\[\baselineskip]
\end{abstract}

\section{\label{sec:intro}Introduction}
Being the simplest nucleus, deuteron provides the most direct test of the various nucleon-nucleon interaction models and the relevant degrees of freedom.
In this context, the deuteron electromagnetic studies through electron or photon probes are the simplest in theoretical and experimental aspects.
These studies provide picture of deuteron electromagnetic structure in terms of deuteron electromagnetic (EM) form factors (FFs).
These FFs depend on the square of the four-momentum transferred by a probe ($q^2=-Q^2$).

During the last two decades  a considerable advance has been made in the experimental knowledge of deuteron electromagnetic structure.
On the other hand there is a substantial diversity of opinion
regarding an appropriate theoretical general approach. Though, it seems natural as well as confirmed by the general data analysis \cite{GilmanGross}
that in the  space-like region of
$Q^2$ (corresponding to the elastic scattering) a successful theory may be obtained from a relativistic description of only the $NN$ channel together with minor modifications
of the short-range structure of the deuteron electromagnetic current (EMC) operator. Here again, there are different approaches concerning the
relativistic description as well as the EMC operator structure \cite{GilmanGross,Garson}.

It is customary to assume that most of the existing data of $eD$ elastic scattering are described to high precision in the one-photon exchange approximation
and by three electromagnetic deuteron FFs \cite{GilmanGross,GilGr25,Donnelly,Akhiezer}. The deuteron FFs are
calculated from the deuteron wave function ($S$ and $D$ components) and from the nucleon FFs in the conventional model.
 FFs may be chosen so as to be equal at $Q^2=0$ (static) limit to the deuteron charge, magnetic and quadruple momenta.
 The first two are described by the conventional nuclear model with only nucleon degrees of freedom.
But deuteron static electric quadrupole moment is not reproduced well enough
by the modern $NN$ potential calculations.
It is generally agreed that the meson exchange contributions must be taken into account for agreement with data.

The one-photon exchange approximation assumes that the electron and deuteron exchange a single virtual  photon. It is believed that
for the most part this approximation must be valid to high precision because of the small value of the  fine-structure constant.
Therefore the elastic $eD$ scattering allows to extract the deuteron EM FFs dependencies on the transferred 4-momentum $Q$ in
the space-like region. To extract these dependencies it is required to measure three independent observables of the $eD$ elastic
scattering in the region. Two of them (structure functions A and B) are extracted from  the unpolarized differential cross section
and third one is extracted from polarization measurements.

At low $Q\lesssim 0.7\ GeV/c \approx 3.5 \ Fm^{-1}$ simple non relativistic calculations with realistic $NN$ potentials (with $NN$ channel
only) agree well enough with one another and with the $eD$ elastic scattering data \cite{GilmanGross,Garson}. With rise of $Q>3.5 \ Fm^{-1}$
disagreement increases between various calculations. None of the calculations describes data for $Q\gtrsim 3.5 \ Fm^{-1}$.
This disagreement indicates that relativistic effects and effects of other channels may be essential at $Q>3.5 \ Fm^{-1}$   \cite{GilmanGross,Garson}.
Indeed, inclusion of relativistic and meson exchange corrections gives good description of  the data \cite{ArenRitz,Schiavilla_2002}.
There is another problem here that is not usually emphasized. The important ingredients of the calculations are nucleons FFs dependencies
on the $Q_{i}^{2}$ transferred to the individual nucleon.
These dependencies are extracted experimentally. The proton FFs are extracted from direct measurements with proton target.
Nevertheless, even in this case there is a notable discrepancy between the extracted values of the proton FF ratio, $G_{Ep}/G_{Mp}$,
from polarization and those obtained from cross section experiments. The cross sections are necessary to extract the absolute value of $G_{Ep}$ and $G_{Mp}$,
while polarization transfer measurements give only ration $G_{Ep}/G_{Mp}$. The discrepancy begins at $Q \approx 1 \ GeV/c \approx 5 \ Fm^{-1}$.
It may be explained by the hard two-photon exchange process (TPE) and data show some evidences of the explanation \cite{Punjabi2015}.
It also should be noted that model calculations and analyses show that TPE significantly change values of $G_{Ep}$, while $G_{Mp}$ changes
at the few percent level ($\sim 3\%$) \cite{Brash2002, Arrington2003}. 
 The latest analytical fit for the proton FFs is a simultaneous fit of the polarization and
cross section data. The cross section data are corrected by an additive term assuming some phenomenological expressions of the TPE correction \cite{Alberico_2009}.

A free neutron target does not exist. The neutron FFs
are extracted from measurements of $eD$ or $e\, ^3 He$ scattering.
Therefore, the data analysis is affected by uncertainties
stemming from the  assumed nuclear theoretical model to describe
the target nucleus and the used reaction.

A common procedure of neutron FFs extraction may be exemplified as follows.  In \cite{Rioradan_2010} the electric FF of the neuteron was measured up to $Q^2=3.4 GeV^2/c^2$ using $^3 \vec{He}(\vec{ e},e'n)pp$ reaction. Details of extraction include
calculations of the asymmetries in the quasi-elastic processes
$^3 \vec{He}(\vec{ e},e'n)pp$ and $^3 \vec{He}(\vec{ e},e'p)np$. These calculations
were performed using the generalized eikonal approximation (GEA), and included the spin-dependent final-state interactions and meson-exchange currents, and used
the $^{3}He$ wave function that results from the AV18 potential. Finally, to extract $G_{En}$ the
linearly interpolated values of $G_{Mn}$ from \cite{Lachniet_2009} were used. Procedure of \cite{Lachniet_2009} is a measurement of
the ratio R of the cross sections for the $^{2}H(e, e′n)p$ and
$^{2}H(e, e′p)n$ reactions in quasielastic scattering on
deuterium. To extract $G_{Mn}$ from the R they use:
1) The cross section calculation using the Plane Wave
Impulse Approximation (PWIA) for $Q^2 >1.0\ GeV^2/c^2$, the AV18 deuteron wave function,
and Glauber theory for final-state interactions (FSI);
2) Calculation of the nuclear correction factor being the ratio
of the full calculation to the PWIA without FSI; 3) The proton
cross section (parametrization of \cite{Arrington2003});
4) And finally parameterizations of $G_{En}$ \cite{Galster_1971, Lomon_2002}.
Obviously it looks like a vicious circle. There is a model insensitive technique of $G_{En}$ extraction \cite{Aren_quasi} but it is applicable only to
the quasi-free kinematics of the $d(\vec{e},e'\vec{n})p$ reaction.  This technique was used for extraction of $G_{En}$ at $Q^2=0.255\ GeV^2$ ($Q\approx 2.6\ Fm^{-1}$) \cite{Eden_GEn1994}
and gave $G_{En}=0.066\pm 0.036\pm 0.009$.
The ``scale'' uncertainty ($\pm 0.009$) was estimated using an independent measurements of $G_{Mn}$, and
some ``Arenh\"{o}vel's model''.  That sounds convincing and may mean Argonne V14 or other contemporary phenomenological potentials used in the estimation.
But there is no estimation of the model systematic error and the whole procedure is an obvious vicious circle.

There are various relativistic models for calculation of the deuteron EM FFs \cite{ArenRitz,AdamAren,Tamura,AllenKlinkFormF,LevPace}.
All of them are reasonable but they may give different results. That is not a contradiction. The question is discussed in the last part of the paper.
In this paper we extend our previous investigations where we described elastic  $NN$ scattering up to 3~$GeV$ of laboratory energy \cite{NNscatMy3},
and electromagnetic reactions with two nucleons:  bremsstrahlung in the $pp$ scattering  $pp\rightarrow pp\gamma$
\cite{ppgammaMy22}, photodisintegration of deuteron $\gamma D\rightarrow np$ \cite{gammaDMy1,gammaDMy2,gammaDMy3},
exclusive electrodisintegration of deuteron  \cite{ednpMy} and the $eD$ elastic scattering \cite{Me_eD_elastic1,Me_eD_elastic2}. We apply manifestly covariant relativistic quantum
mechanics (RQM) \cite{RKM_Review} in point-form (PF). The PF is one of the three forms of RQM
proposed by Dirac \cite{Dirac}. The other two in common use are the instant form and
front  form. Each form is associated with a subgroup  of the Poincar\'{e} group.
 This subgroup is considered to be free of interactions.
All the forms are unitary equivalent \cite{Sokolov1984}, however each form has certain advantages. In particular, the PF has some simplifying features \cite{Klink}.
Only in the PF  all generators of the homogeneous Lorentz group are free of interactions. That means manifest covariance, and clearly simplifies boost transformations.
Therefore the spectator  approximation (SA) of an electromagnetic process preserves its
spectator character in any frame of reference (f.~r.) \cite{Lev3,Melde,Desplanques}.
There are two equivalent  SAs of EM current operator for composite systems in PF RQM  \cite{Klink,Lev3}. The PF RQM SA was applied to calculate EM FFs
 of composite systems \cite{AllenKlinkFormF,PFSA1,PFSA2,Wagen,Amghar,CoesterFF2} with satisfactory results.

\section{Potential model in PF of RQM}
General method for putting interactions in
generators of the Poincar\'{e} group was
proposed by Bakamjian and Thomas \cite{BakamjianThomas}.
We give only results of PF RQM  necessary
for our $eD$ calculation. We use
formalism and notation of \cite{Lev3} for calculation of the matrix elements of
the EM current operator.


Let $p_i$ be the 4-momentum  of nucleon
$i$, $P\equiv(P^0,{\bf P})=p_1+p_2$ be the system 4-momentum, $M$ be
the system mass and $G=P/M$ be the system 4-velocity. The system wave
function  with 4-momentum $P$ is expressed through a
tensor product of external and internal parts
\begin{equation}
\vert P,\chi\rangle
=U_{12}\,\vert P\rangle\otimes\vert \chi\rangle,
\end{equation}
where the internal wave function $\vert\chi\rangle$ satisfies
Eqs.~(\ref{rel1})-(\ref{rel2}). The unitary operator
\begin{equation}
U_{12}=U_{12}(G,{\bf q})
=\prod_{i=1}^{2} D[{\bf
s}_i;\alpha(p_i/m)^{-1}\alpha(G)\alpha(q_i/m)] \label{U12}
\end{equation}
is the  operator from the ''internal'' Hilbert space  to the
Hilbert representation space of two-particle states \cite{Lev3}.
$D[{\bf s};u]$ is the SU(2) representation (matrix) operator
corresponding to the element $u\in$SU(2).  ${\bf s}$ are generators of the  representation.
In case of spin $s$ = 1/2 particles, we deal with the
fundamental representation. In this case ${\bf s}_i\equiv
\frac{1}{2}{\bm \sigma}_i$ (${\bm \sigma} =( \sigma_x,
\sigma_y, \sigma_z)$ are the Pauli matrices) and $
D[ {\bf s}; u]\equiv u$.
The momenta of the particles in their c.m. frame are
\begin{equation}
q_i=L[\alpha(G)]^{-1}p_i, \label{scm_q}
\end{equation}
where $L[\alpha(G)]$ is the Lorentz transformation to the frame
moving with 4-velocity $G$. Matrix
\begin{equation}
\alpha (g)= {(g^0+1+ {\bf  \sigma }\cdot {\bf g)}}/{
\sqrt{2(g^0+1)}}
\end{equation}
 corresponds to a 4-velocity $g\equiv \left(g^{0},{\bf g}\right)$.

The external part of the wave function is defined as
\begin{gather}
\label{ext_wf}  \langle G\vert
P'\rangle\equiv\frac{2}{M'}G^{'0}\delta^3({\bf G}-{\bf G}').
\end{gather}
Its scalar product is
\begin{equation}
\label{ext_wf_product}  \langle P''\vert P'\rangle=\int
\frac{d^3{\bf G}}{2G^0} \langle P''\vert G\rangle \langle G\vert
P'\rangle
=2\sqrt{M^{'2}+{\bf P}^{'2}}\delta^3({\bf P''}-{\bf P}'),
\end{equation}
where $G^{0}({\bf G})\equiv\sqrt{1+{\bf G}^2}$.
The internal part
 $\vert\chi\rangle$ is characterized by the system full angular momentum $J$ and by momentum  ${\bf q}={\bf q}_1=-{\bf q}_2$ of one of the particles in the
 c.m. frame.

 Interaction appears in 4-momentum $\hat{P}=\hat{G}\hat{M}$, where $\hat{M}$ is sum of the free mass
operator ${M}_{free}$ and of the interaction $V_{int}$: $\hat{M}={M_{free}}+V_{int}$.
 The wave function is an eigenfunction of the system mass operator
$\hat{M}$. We represent this wave function as a
product of the external and internal parts. The internal
wave function $\vert\chi\rangle$ is also an eigenfunction of the
mass operator and for the system of two nucleons with masses
$m_{1}\approx m_{2}\approx m=2m_{1}m_{2}/(m_{1}+m_{2})$ satisfies the following equation
%
\begin{equation}
\label{rel1} \hat{M}\vert\chi\rangle\equiv \left[ 2\sqrt{{{\bf
q}}^2+m^2}+V_{int} \right]\vert\chi\rangle = M\vert\chi\rangle.
\end{equation}
The Eq.~(\ref{rel1}) may be rearranged as
\begin{equation}
\label{rel2} \left[ {{{\bf q}}^2 + m V} \right]\vert\chi\rangle = q^2\vert\chi\rangle,
\end{equation}
where $V$ acts only through internal variables and
\begin{equation}
  q^2 = \frac{M^2}{4} - m^2.\label{q2}
\end{equation}
The interaction operator acts only
through internal variables. Operators  $V_{int}$
and $V$ (and therefore  $\hat{M}$ and  ${M}_{free}$) commute with spin operator $J$  and
with 4-velocity operator $\hat{G}$. Generators of space-time rotations  are free of interaction.
Most non-relativistic scattering theory  formal results are valid
for  case of two particles \cite{RKM_Review}. For example in the c.m. frame
the relative orbital angular momentum and spins are
 coupled together as in the non-relativistic case.
Eq.~(\ref{rel2}) is identical in form to the Schr\^{o}dinger
equation. The only relativistic correction here is in the deuteron binding energy
that must be changed  by the effective value $2.2233$ MeV instead of the experimental $2.2246$ MeV.
It is easy to show. Let $\varepsilon$ be the deuteron binding energy.
Then $M=2m-\varepsilon$ and for deuteron state of the $NN$ system  $q^2=\frac{M^2}{4}-m^2=-m\varepsilon \left(1-\frac{\varepsilon}{4m}\right)$.
Comparing with the nonrelativistic relationship $q^2=-m\varepsilon$
we identify factor $\left(1-\frac{\varepsilon}{4m}\right)$ as the relativistic correction.
There is no similar correction in the scattering region because $q^2=mE_{lab}/2$ is the precise relativistic relationship ($E_{lab}$ is the laboratory energy), that is used in the partial wave analysis.
This correction is negligible for our problem.

The deuteron wave function $\vert P_{i},\chi_{i}\rangle$ is
normalized as follows
\begin{equation}
\label{sc_prod_deut} \langle P_f,\chi_{f}\vert
P_i,\chi_{i}\rangle=2 P_{i}^{0}\,\delta^3({\bf P}_i-{\bf
P}_f)\langle \chi_{f}\vert
\chi_{i}\rangle.
\end{equation}


\section{$eD$ elastic scattering}
   There is a convenient r.~f. for calculation of current operator matrix elements in PF of RQM introduced by F. Lev \cite{Lev3} (it coincides for  elastic $ed$ scattering with the Breit r.~f.).
   For all EM reactions with two nucleons this Lev r.~f. is defined  by condition:
\begin{equation}
{\bf G}_{f}+{\bf G}_{i}=0,
 \label{cond1}
\end{equation}
where ${\bf G}_f={\bf P}_f/M_D$, ${\bf G}_i={\bf P}_i/M_D$ are final and initial 4-velocities of the deuteron and $M_D$ is its mass.
 The matrix element of the current operator is  \cite{Lev3}:
\begin{equation}
\langle P_{f},\chi_{f}\vert \hat{J}^{\mu}(x)\vert P_{i},\chi_{i}\rangle
=2(M_{f}M_{i})^{1/2} \exp(\imath (P_f - P_i)x)\langle\chi_f\vert \hat{j}^{\mu}({\bf h})\vert\chi_i\rangle , \label{mat1}
\end{equation}
where the internal current operator $\hat{j}^{\mu}({\bf h})$ defines action of current operator in the internal space of the $NN$ system.
\begin{equation}
{\bf h}=\frac{2(M_i M_f)^{1/2}}{(M_i+M_f)^{2}}\,{\bf k}=\frac{\bf k}{2M_{D}}
\end{equation}
is vector-parameter \cite{Lev3} ($0\leq h \leq 1$), ${\bf k}$ is momentum of photon in r.~f.  (\ref{cond1}), $M_{i}=M_{f}=M_{D}$ are masses of initial and final $NN$ system (deuteron).

The internal wave function of deuteron  is
\begin{gather}
\vert\chi_i\rangle=
\frac{1}{r}\sum_{l=0,2}u_{l}(r)\vert l,1;J=1M_{J} \rangle_{{\bf r}} \label{wf_D},
\end{gather}
normalised as  $\langle \chi_i\vert\chi_i\rangle=1$. This configuration space wave function may have physical sense only in nonrelativistic limit.
In our calculations we use the momentum space wave function:
\begin{gather}
\vert\chi_i\rangle=\frac{1}{q}\sum_{l=0,2}u_{l}(q)\vert l,1;1M_{J} \rangle_{{\bf q}},
\end{gather}
where
\begin{gather}
u(q)\equiv u_{0}(q)=\sqrt{\frac{2}{\pi}}\int dr \sin(qr) u(r),\label{MSpaceU}
\end{gather}
\begin{equation}
w(q)\equiv u_{2}(q)=\sqrt{\frac{2}{\pi}}\int dr \left[ \left(\frac{3}{(qr)^2}-1\right)\sin(qr)\right.\label{MSpaceW}
-\left.\frac{3}{qr}\cos(qr)\right]w(r).
\end{equation}

Transformations from the Breit r.~f. (\ref{cond1}) to the final c.~m. frame of the $NN$ system and  to the initial one are boosts along and against vector
 ${\bf h}$ (axis $z$) correspondingly. Projection of the total deuteron angular  momentum onto the  $z$ axis does not change for these boosts.
The initial deuteron in the Breit r.~f.  moves in direction opposite  to the ${\bf h}$. Its internal  wave function with spirality   $\Lambda_i$
  is
\begin{equation}
\vert \Lambda_i\rangle=\frac{1}{q}\sum_{l=0,2}u_{l}(q)\vert
l,1;1,M_{J}=-\Lambda_i \rangle. \label{wi_D_spir}
\end{equation}
 Wave function of the final deuteron with spirality   $\Lambda_f$ is
\begin{equation}
\vert \Lambda_f\rangle =\frac{1}{q}\sum_{l=0,2}u_{l}(q)\vert
l,1;1,M_{J}=\Lambda_f \rangle, \label{wf_D_spir}
\end{equation}

Usual parametrisation of the EM CO matrix element for a spin-1 particle (deuteron) is  \cite{Arnold81,GilmanGross,Garson}:
\begin{multline}
(4P_i^{0}P_f^{0})^{1/2}\langle P_f,\chi_f|J^{\mu} |P_i,\chi_i\rangle\\
=- \left\{ G_1(Q^2)({\bm\xi}_f^{*}\cdot {\bm\xi}_i)-G_3(Q^2)\frac{({\bm\xi}_f^{*}\cdot \Delta P)({\bm\xi}_i\cdot \Delta P)}{2M_{D}^2}\right\}(P_i^{\mu}+P_f^{\mu})\\
-G_2(Q^2)[{\xi}_i^{\mu}({\bm\xi}_f^{*}\cdot \Delta\textbf{ P})-{\xi}_f^{*\mu}({\bm\xi}_i\cdot \Delta \textbf{P})],
\end{multline}
where $(a\cdot b)=a^0 b^0 -({\bf a}\cdot {\bf b})$, EM FFs $G_i(Q^2)$, $i=1,2,3$ are function of $Q^2=-\Delta P^2$, $\Delta P=P_f-P_i$.
In the Breit r.~f. $\textbf{P}_f=-\textbf{P}_i$,  $P_i^{0}=P_f^{0}\equiv P^0=M_D /\sqrt{1-h^2}$, ${\Delta P=(0, 2\textbf{P}_f)}$,
${P_i^{\mu}+P_f^{\mu}=(2P^0,\textbf{0})}$, ${\bf P}_f/P^0={\bf h}$, ${\bf P}_f={\bf h}M_D/\sqrt{1-h^2}$, \\ ${\Delta P^2=-4h^{2} M_D^2/(1-h^2)}$, ${Q^2\equiv -\Delta P^2}$,
$h^2=({\bf h}\cdot {\bf h})$. Matrix elements of the internal current operator are
\begin{multline}
\langle \chi_f|j^{0}({\bf h}) |\chi_i\rangle =
 - G_1(Q^2)({\bm\xi}'^{*}\cdot {\bm\xi})+2G_3(Q^2)\frac{({\bm\xi}_{f}^{*}\cdot {\textbf{ h}})({\bm\xi}_i\cdot {\textbf{h}})}{1-h^2}\\
+G_2(Q^2)[{\xi}_i^{0}({\bm\xi}_{f}^{*}\cdot {\bf h})-{\xi}_f^{0*}({\bm\xi}_i\cdot  \textbf{h})],
\end{multline}
\begin{equation}
\langle \chi_f|{\bf j}({\bf h}) |\chi_i\rangle =
G_2(Q^2)[\xi_i({\bm\xi}_f^{*}\cdot {\bf h})-\xi_f^{*}({\bm\xi}_i\cdot {\bf h})]=
G_2(Q^2)[{\bf h}\times [{\bm\xi}_i\times {\bm\xi}_f^{*}]].
\end{equation}

%
It can be shown  \cite{Lev3}, that these expressions are equivalent to $j^{\nu}$ defined as:
\begin{align}
j^0({\bf h})&=
G_C(Q^2)+\frac{2G_Q(Q^2)}{(1-{h}^2)}
\left[\frac{2}{3}{ h}^2-({\bf h}\cdot {\bf J})^2\right],\label{123}\\
{\bf j}({\bf h})&=-\frac{\imath }{\sqrt{1-{h}^2}}
G_M(Q^2)({\bf h}\times {\bf J}),
\end{align}
where ${\bf J}$ is the total angular momentum (spin) of the deuteron. $G_C$, $G_Q$, $G_M$ are its charge monopole, charge quadruple and magnetic dipole FFs.

Spiral polarizations  of the deuteron in the initial and final states are
\begin{align}
\xi_{i}^{\Lambda}&=\left\{
\begin{array}{l@{{}\ \ \ \ \ \ \ \ {}}r}
(0,\pm 1,-\imath,0)/\sqrt{2}& (\Lambda=\pm) \\
(-h,0,0,1)/\sqrt{1-h^2}& (\Lambda=0),\\
\end{array}
\right.\\
 \xi_{f}^{\Lambda}&=\left\{
\begin{array}{l@{{}\ \ \ \ \ \ \ \ \ \ \ \ \ {}}r}
(0,\mp 1,-\imath,0)/\sqrt{2}& (\Lambda=\pm) \\
(h,0,0,1)/\sqrt{1-h^2}&(\Lambda=0).\\
\end{array}
\right.
\end{align}

Polarization of the virtual photon is
\begin{equation}
\epsilon^{\lambda}=\left\{
\begin{array}{l@{{}\ \ \ \ \ \ \ \ {}}r}
(0,\mp 1,-\imath,0)/\sqrt{2}& (\lambda=\pm) \\
(1,0,0,0)& (\lambda=0).\\
\end{array}
\right.
\end{equation}

The deuteron FFs $G_i$ are expressed as
\begin{equation}
\begin{split}
G_C&=G_1+\frac{2}{3}\eta G_{Q},\\
G_{Q}&=G_1-G_M+(1+\eta)G_3,\\
G_1&=G_C-\frac{2h^2}{3(1-h^2)}G_Q,\\
G_3&=G_Q\left(1-\frac{h^2}{3}\right)-G_C(1-h^2)+G_M(1-h^2),
\end{split}
\end{equation}
where $\eta=Q^2/4M_D^2=h^2/(1-h^2)$.
Form factors ${G_C(0)=e}$, $G_M(0)=\mu_{D}e/2 M_D$ and $G_Q(0)=Q_{D}e/M^2_D$ give charge, magnetic and quadruple momenta of deuteron.

Defining helicity amplitudes as  $j^{\lambda}_{\Lambda_{f}\Lambda_{i}} \equiv \langle \Lambda_f | \left( \epsilon^{\lambda}_{\mu}\cdot j^{\mu}({\bf h}) \right) | \Lambda_i \rangle$,
we arrive at:
\begin{equation}
j^{0}_{00}(Q^2)=G_{C}+\frac{4}{3}\frac{h^2}{1-h^2}G_{Q},
\end{equation}
\begin{equation}
j^{0}_{+-}(Q^2)=j^{0}_{-+}(Q^2)=G_{C}-\frac{2}{3}\frac{h^2}{1-h^2}G_{Q},
\end{equation}
\begin{equation}
\frac{j^{+}_{+0}(Q^2)+j^{+}_{0-}(Q^2)}{2}=-\frac{h}{\sqrt{1-h^2}}G_{M}
\end{equation}
and
\begin{equation}
j^{+}_{+0}(Q^2)=j^{-}_{-0}(Q^2)\approx j^{+}_{0-}(Q^2)=j^{-}_{0+}(Q^2).\label{j000}
\end{equation}

The deuteron FFs squared are extracted from the elastic $eD$ scattering with unpolarised particles and an additional polarisation observable (usually $t_{20}(Q^{2},\theta)$).

In the present paper we use SA of the EM CO of \cite{Lev3} without expanding it in powers of $h$ and we calculate its matrix elements in the momentum space.
Therefore calculating
 (\ref{j000}) we use a following precise expansion of  $\hat{j}^{\mu}({\bf h})\approx \hat{j}_{SA}^{\mu}({\bf h})$ \cite{ednpMy}
\begin{multline}
\hat{j}_{SA}^{\mu}({\bf h})=\left(1+({\bf A}_{2}\cdot{\bf s}_2)\right)\left(B^\mu_{1}+({\bf C}^\mu_{1}\cdot{\bf s}_1)\right){\bf I}_{1}({\bf h})\\
+ \left(1+({\bf A}_{1}\cdot{\bf s}_1)\right)\left(B^\mu_{2}+({\bf C}^\mu_{2}\cdot{\bf s}_2)\right){\bf I}_{2}({\bf h}),
\end{multline}
where  ${\bf A}_{i}$, $B^\mu_{i}$, ${\bf C}^\mu_{i}$ are some cumbersome vector functions of  ${\bf h}$ and ${\bf q}(q,\theta,\phi)$.
In the spherical coordinate system $(q,\theta,\phi)$ dependence of these functions on $\phi$ appears as $e^{\pm im\phi}$ ($m=0,1,2$). The $\phi$ is analytically integrated giving trivial equalities in (\ref{j000}).
\section{Results and Conclusions}

\begin{figure}[htb]
\centerline{\includegraphics[width=0.5\textwidth]{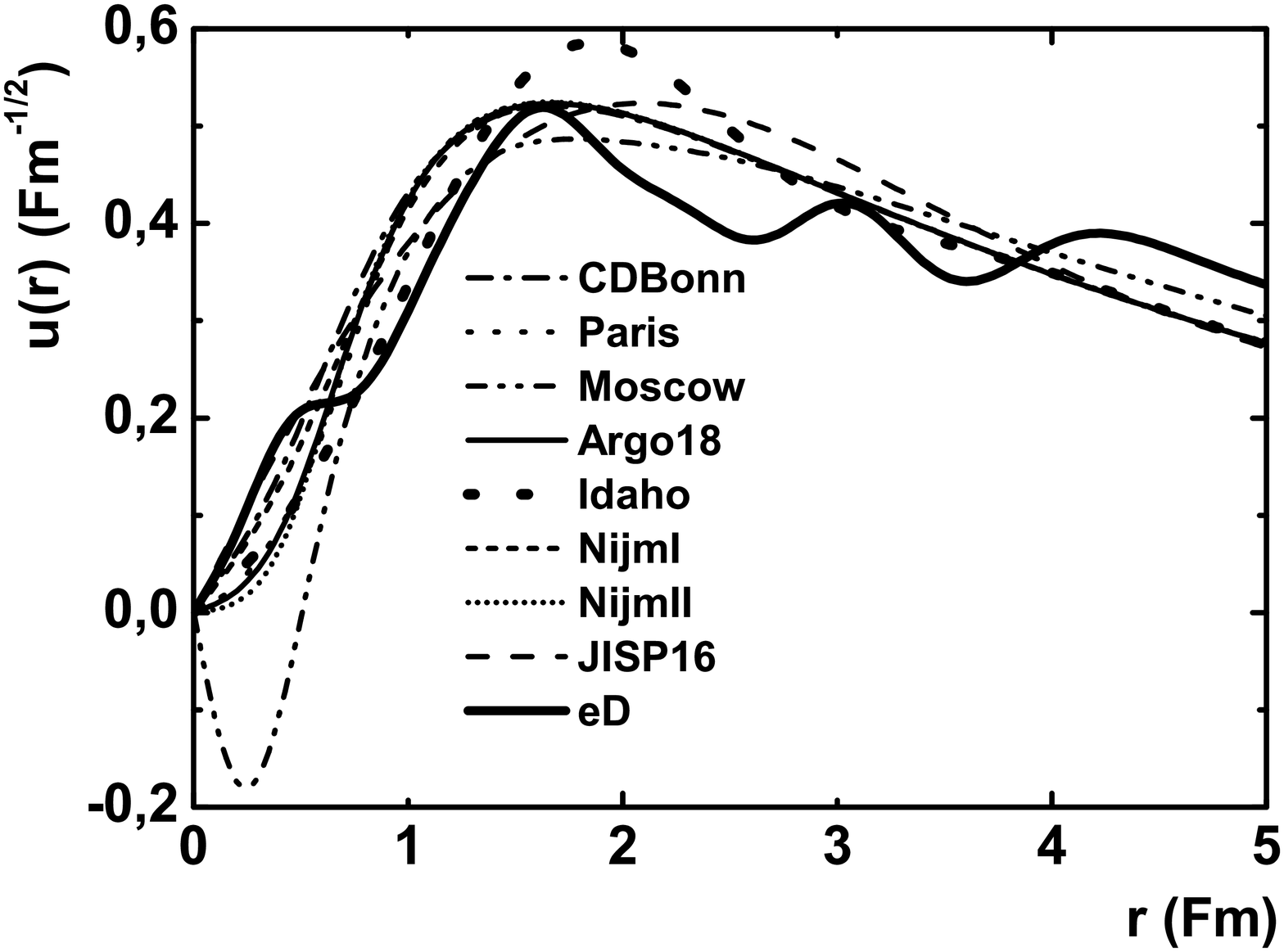}}
\centerline{\includegraphics[width=0.5\textwidth]{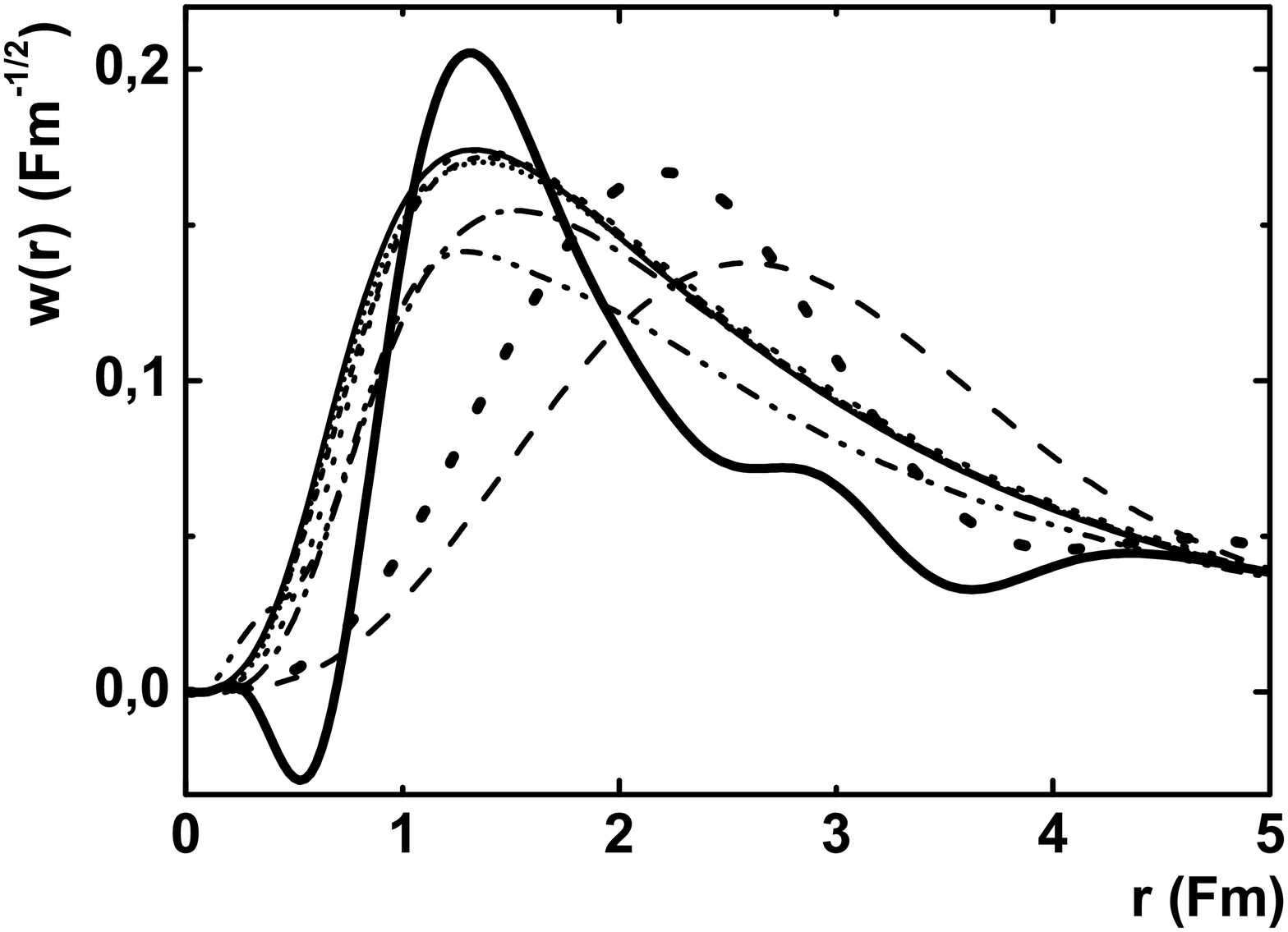}}
\caption{\label{fig:RWs}Configuration space deuteron wave functions  used in the calculations.}
\end{figure}

\begin{figure}[htb]
\centerline{\includegraphics[width=0.55\textwidth]{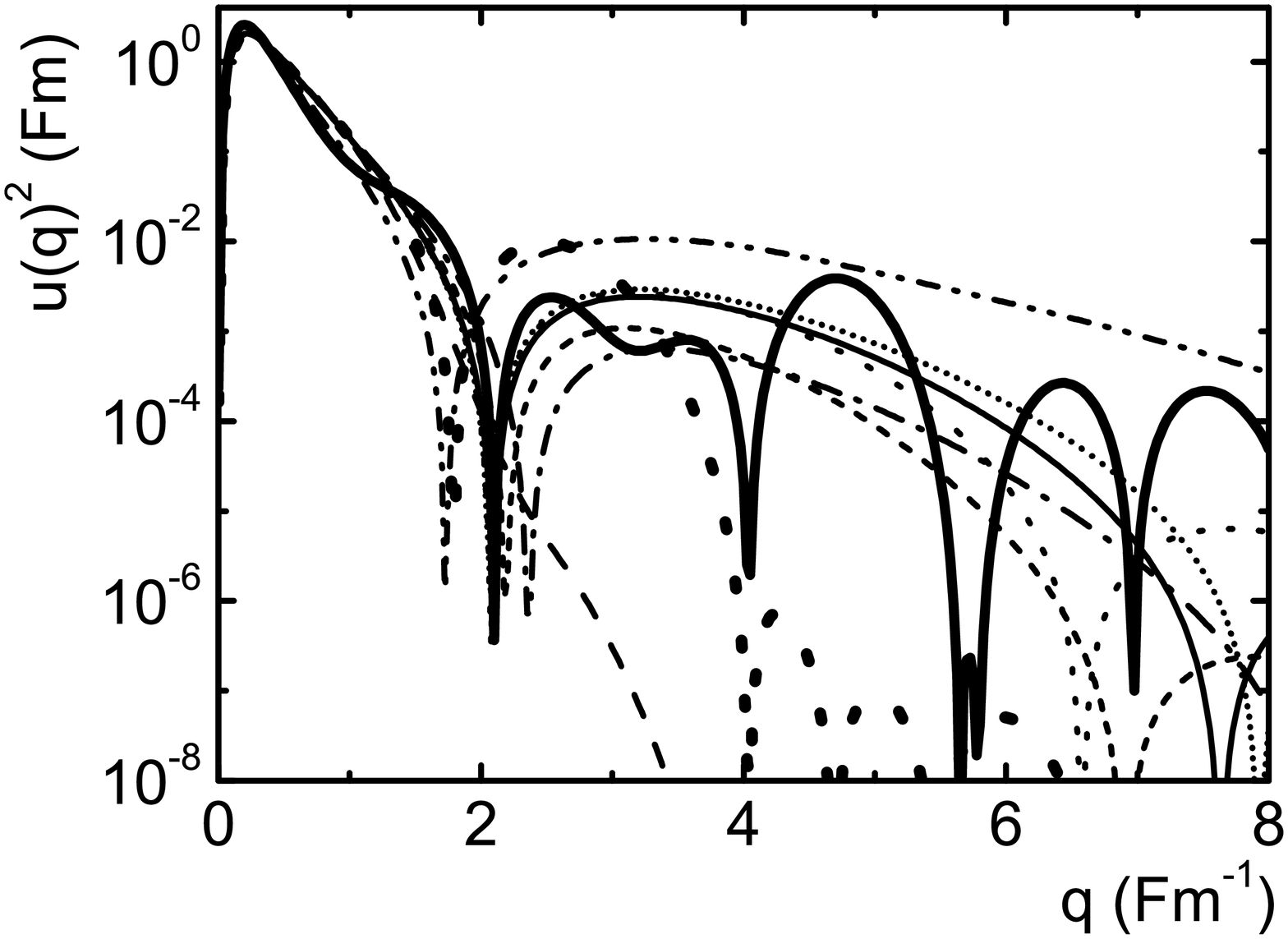}}
\centerline{\includegraphics[width=0.55\textwidth]{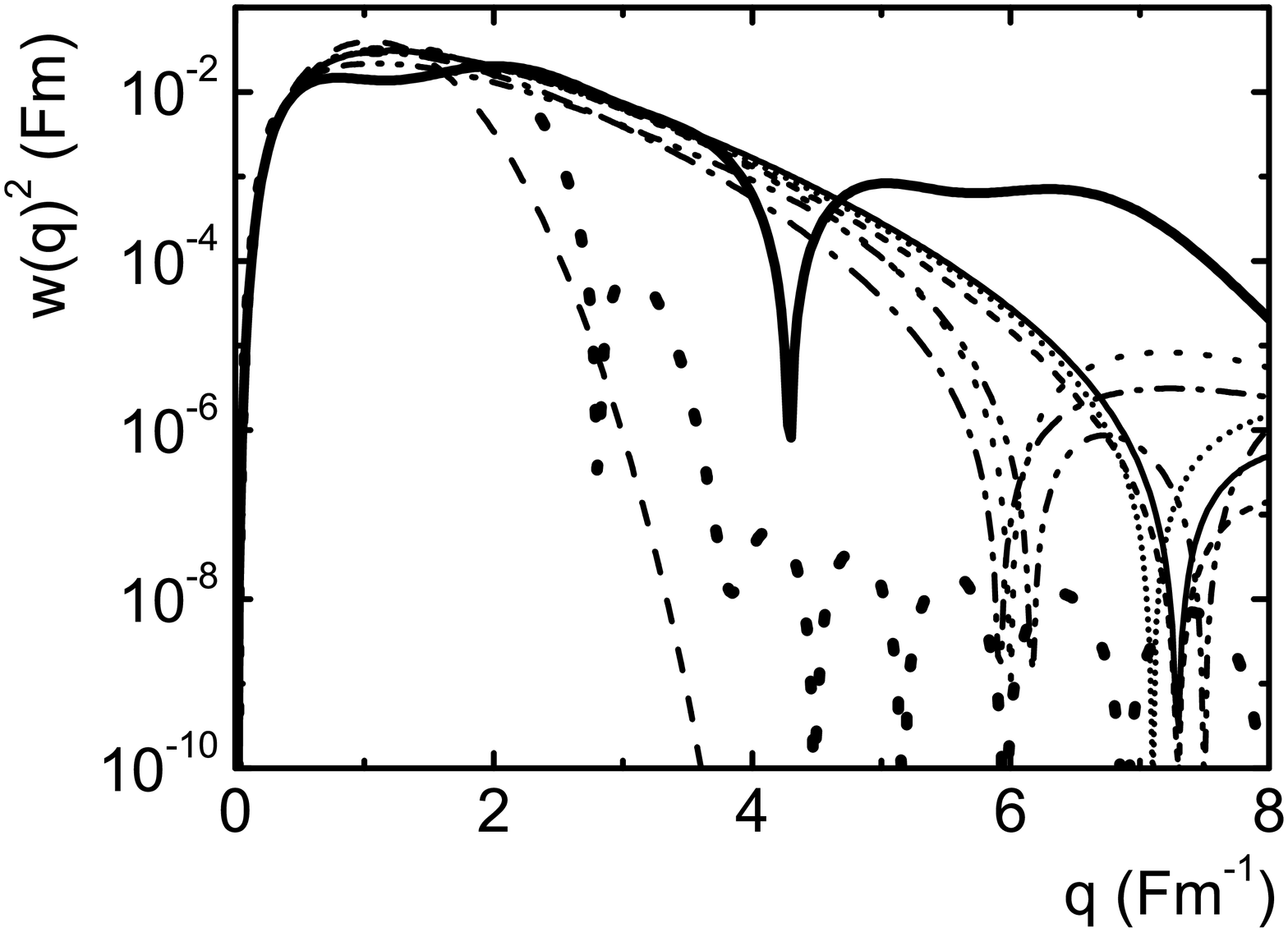}}
\caption{\label{fig:PWs}Momentum space deuteron wave functions  used in the calculations. Curve legends as Fig.~\ref{fig:RWs}.}
\end{figure}

In our investigation we try to describe the $eD$ elastic scattering data by the simplest model. The model is the $NN$ channel described in PF RQM and SA of the $NN$ EMC operator.
Therefore, we assume that effect of the
exchange currents
may be negligible  for some deuteron $NN$ model at least for this reaction.
The momentum space deuteron wave functions may be transformed into the configuration space according to (\ref{MSpaceU}-\ref{MSpaceW}). We assume that configuration wave functions in (\ref{wf_D}) may have
physical sense only in the nonrelativistic limit.
The deuteron wave functions stemming from Nijmegen-I (NijmI), Nijmegen-I (NijmII) \cite{Nijm}, JISP16 \cite{JISP16}, CD-Bonn \cite{CD-Bonn}, Paris \cite{Paris},
Argonne18 \cite{Argonne18} (momentum space deuteron wave function is a parametrisation from \cite{Argonne18ms}), Idaho \cite{Idaho}
(thanks to Prof. David R. Entem for the sent computer code)
and Moscow (with forbidden states) \cite{NNscatMy3} potentials are shown in Figs.~\ref{fig:RWs} and~\ref{fig:PWs}.
Parameters and computer code of the Moscow potentials may be requested from the author (e-mail: nikolakhokhlov@yandex.ru).

In our previous investigations \cite{Me_eD_elastic1,Me_eD_elastic2} we considered dependence of deuteron FFs on the nucleons FFs and we found this dependence considerable at $Q>5\ Fm^{-1}$.
Now we investigate a possibility of extraction of the deuteron wave function and neutron EM FFs from the elastic
$eD$ scattering. Therefore we performed a fitting procedure for the deuteron wave function and nucleons EM FFs.
Functional dependency on $Q^2$ of the nucleon FFs is as in \cite{Arrington2003}, but parameters were fitted. The deuteron wave function parametrization is described bellow.
Results of the fitting procedure are denoted as $eD$ in Table~\ref{tab1} and in all figures.


We use the deuteron wave functions in the  analytic form of \cite{Paris} modified by a short range addend.
The ansatz for the r-space wave functions is
\begin{equation}
u(r)=\sum_{i=0}^{m}a_{i}^{0}R_{i,0}(r)+\sum_{j=1}^{n}C_{j}exp(-m_{j}r),
\end{equation}
\begin{equation}
w(r)=\sum_{i=0}^{k}a_{i}^{2}R_{i,2}(r)
+\sum_{j=1}^{n}D_{j}exp(-m_{j}r)\left[1+\frac{3}{m_{j}r}+\frac{3}{\left( m_{j}r \right)}\right],
\end{equation}
where the oscillator functions
\begin{equation}
R_{i,l}(r)=(-1)^{n}\sqrt{\frac{2n!}{r_{0}\Gamma(n+l+3/2)}}\left(\frac{r}{r_{0}}\right)^{l+1}\times%
exp\left(-\frac{r^{2}}{2r_{0}^{2}}\right)
L_{i}^{l+\frac{1}{2}}\left(\frac{r^{2}}{r_{0}^{2}}\right),
\end{equation}
$L_{i}^{\alpha}$ is the associated Laguerre polynomial, the oscillator radius $r_{0}=0.4\ fm$.

The corresponding momentum space wave functions can be easily obtained analytically by the Fourier transforms.
The boundary conditions at $r=0$ lead to the constraints
\begin{equation}
C_{n}=-\sum_{j=1}^{n-1}C_{j},
\end{equation}
\begin{multline}
D_{n-2}=\frac{m_{n-2}^{2}}{(m_{n}^{2}-m_{n-2}^{2})(m_{n-1}^{2}-m_{n-2}^{2})}\\
\times\left[-m_{n-1}^{2}m_{n}^{2}\sum_{j=1}^{n-3}
\frac{D_{j}}{m_{j}^{2}}+(m_{n-1}^{2}+m_{n}^{2})\sum_{j=1}^{j=n-3}D_{j}-\sum_{j=1}^{n-3}D_{j}m_{j}^{2}\right],
\end{multline}
and two relations obtained by circular permutation of $n-2,n-1,n$. All parameters may be requested from the author in the Fortran code.
\begin{table}[htb]
\caption{Static deuteron form factors.
Two values through slash are relativistic calculation/nonrelativistic calculation.
There are results of relativistic calculation only  in the last three rows.}
\label{tab1}
  \begin{center}
\begin{tabular}{ccc}
\hline\noalign{\smallskip}
 & $G_{M}(0)=\frac{M_{d}}{m_{p}}\mu_{d}$ & $G_{Q}(0)=M_{d}^{2}Q_{d}$  \\
\noalign{\smallskip}\hline\noalign{\smallskip}
Exp & 1.7148 & 25.83 \\
NijmI & 1.697/1.695 & 24.8/24.6 \\
NijmII & 1.700/1.695 & 24.7/24.5 \\
Paris & 1.696/1.694 & 25.6/25.2 \\
CD-Bonn & 1.708/1.704 & 24.8/24.4 \\
Argonne18 & 1.696/1.694 & 24.7/24.4 \\
JISP16 & 1.720/1.714 & 26.3/26.1 \\
Moscow06 & 1.711/1.699 & 24.5/24.2 \\
Moscow14 & 1.716/1.700 & 26.0/25.8 \\
Idaho  & 1.714/1.700   & 26.22/25.98  \\
eD & 1.715/1.700 & 25.83/25.54 \\
\noalign{\smallskip}\hline
\end{tabular}
\end{center}
\end{table}
\begin{figure}[ht]
\centerline{\includegraphics[width=0.45\textwidth]{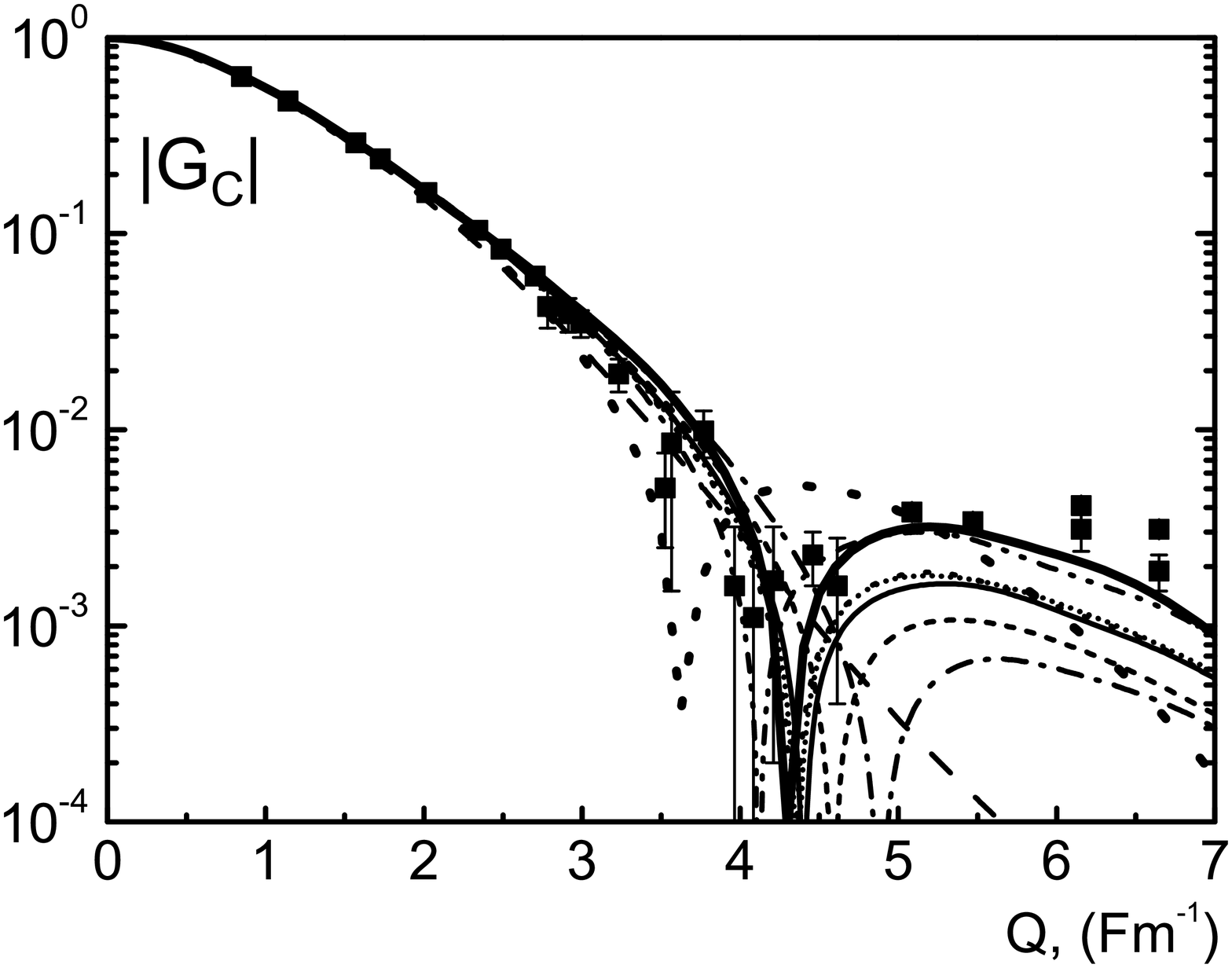}}
\centerline{\includegraphics[width=0.45\textwidth]{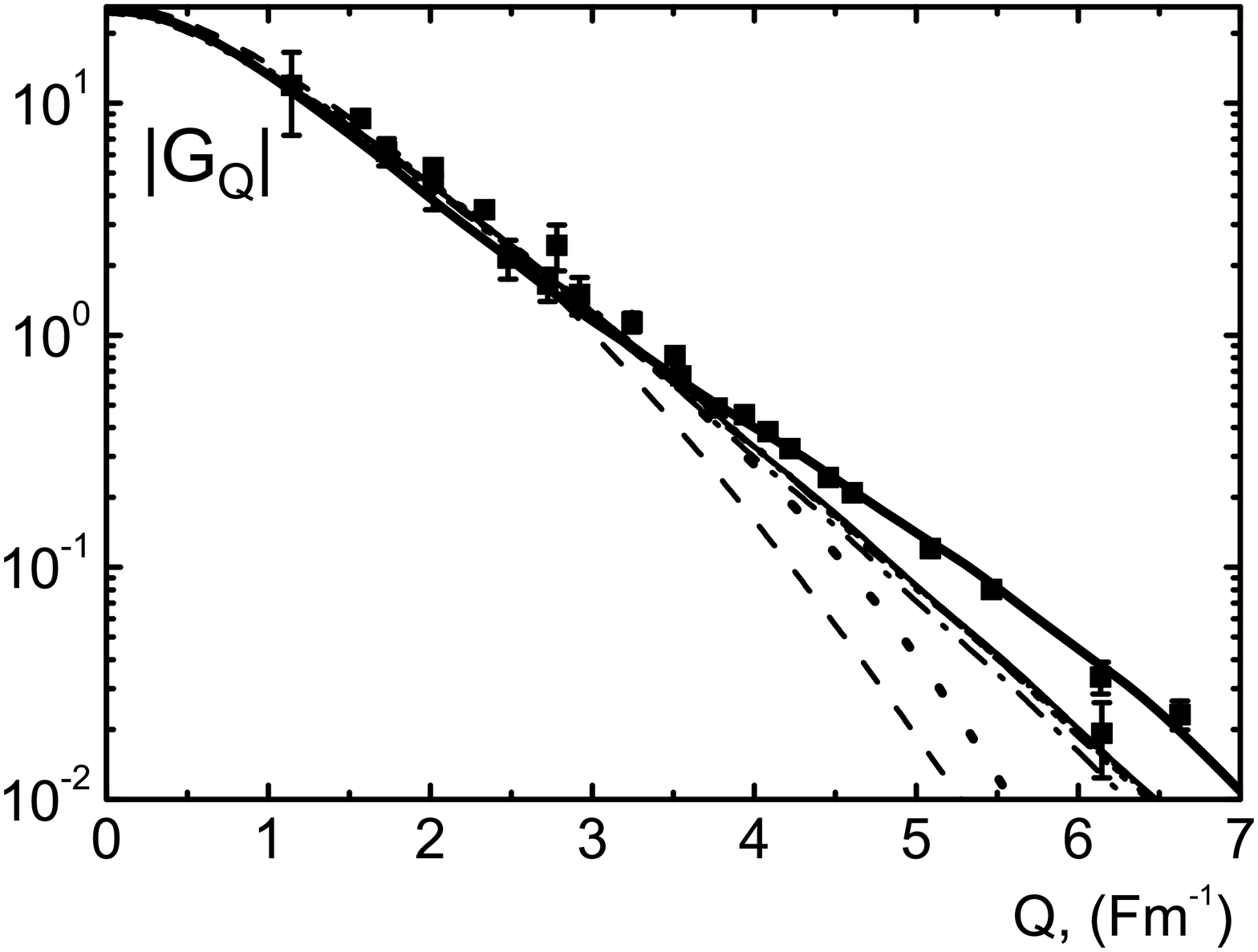}}
\centerline{\includegraphics[width=0.45\textwidth]{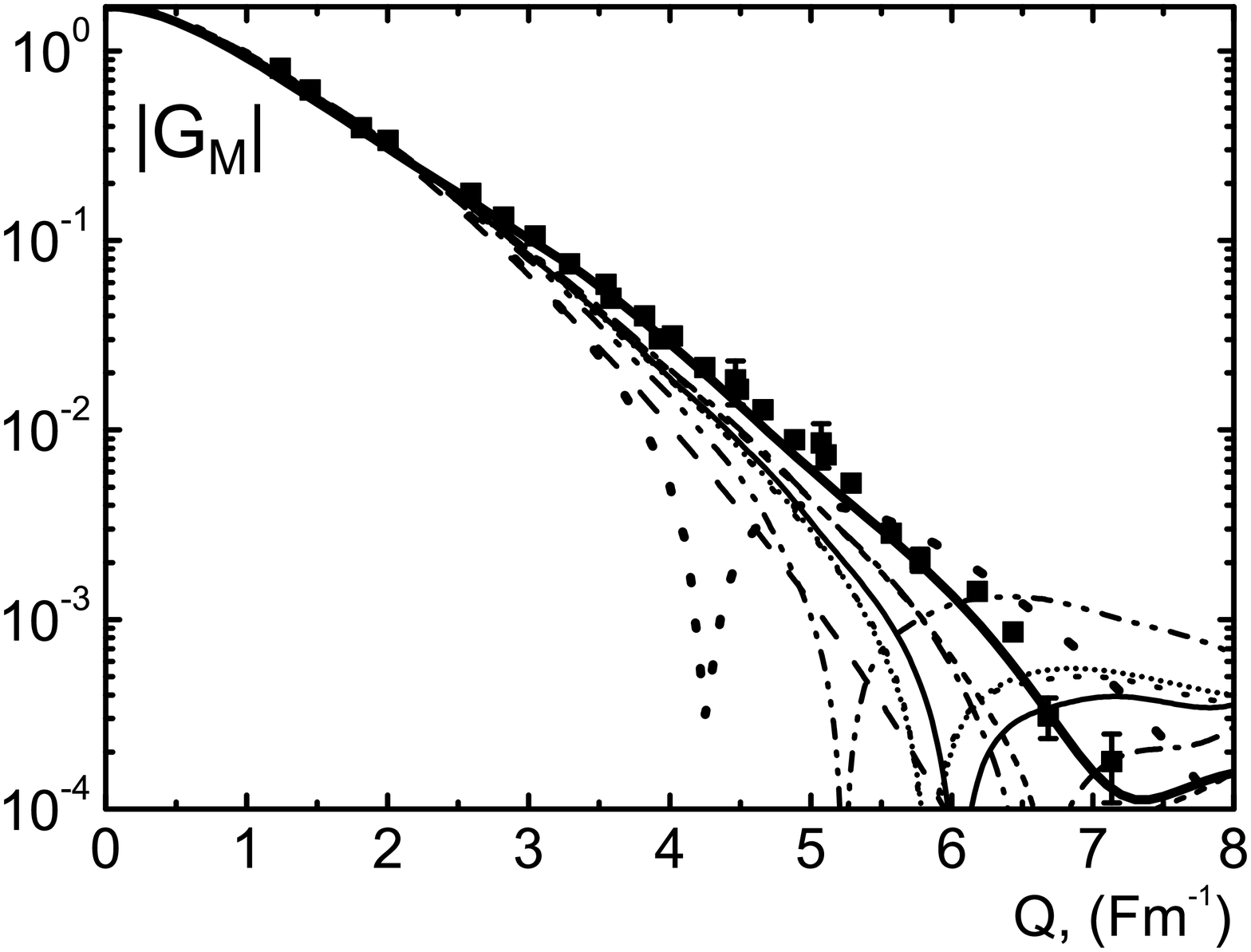}}
\caption{\label{fig:dFFs} Deuteron form factors as a function of Q. Data are from compilation of \cite{Garson}
calculated from $A$, $B$ and $t_{20}$ data of
\cite{r87,r90,r94,r97,r98,r99,r95,r96,r76,r77,r103,r104,r105,r106,r107,r108,r78}. Curve legends as Fig.~\ref{fig:RWs}.}
\end{figure}
\begin{figure}[htb]
\centerline{\includegraphics[width=0.45\textwidth]{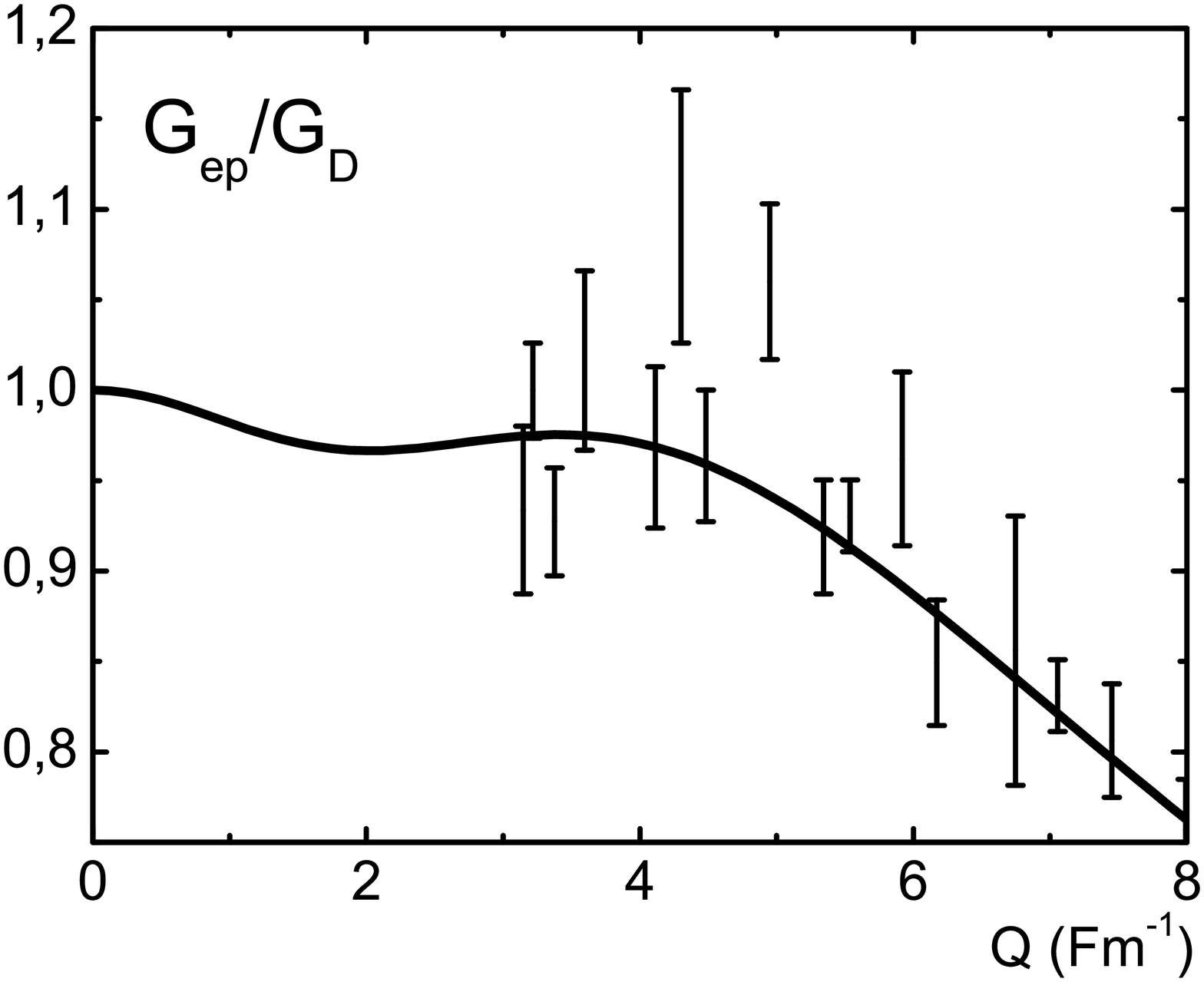}}
\centerline{\includegraphics[width=0.45\textwidth]{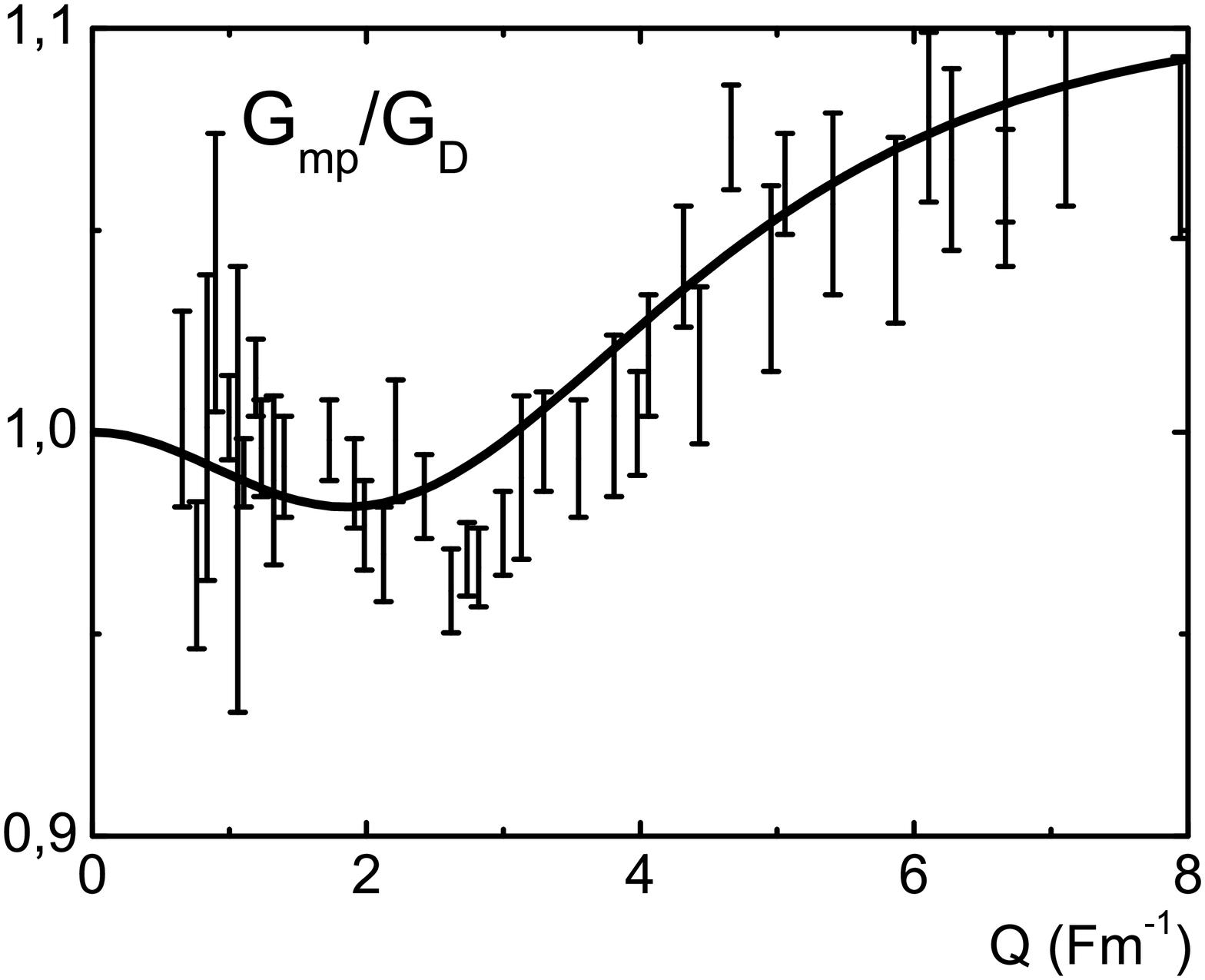}}
\caption{\label{fig:pFFs} Proton EM FFs as a functions of Q. Data are compilation of \cite{MichaelKohl2008} of polarization experiments analysis.}
\end{figure}

Results of our calculations are shown in Figs. 1-5. We see very good general correspondence of the theory and
experiment for $Q< 3$ Fm$^{-1}$. Discrepancies for larger $Q$ are comparable with differences of results for different models (potentials).
Some model calculations  \cite{arenhovel_excange_curr} show that meson exchange currents may give significant effect in EM processes with  $np$-system.
In our calculation we do not take into account these currents. It is not clear how these currents may be agreed with the short range part of the $NN$ interaction of quantum chromodynamic origin. We have are number of deuteron models (Figs. 1 and 2) stemming from
equally justified modern theories  (Argonne18, Idaho) that obviously require different meson exchange currents.
Besides the EM FFs of nucleons are not described by meson degrees of freedom at intermediate and high energies  \cite{Perdrisat2007},
moreover the neutron EM FFs are not measured. As discussed in the Introduction all data of the  neutron EM FFs are model dependent.
Any conclusions on meson exchange currents are premature without solid data of the  neutron EM FFs.
\begin{figure}[htb]
\centerline{\includegraphics[width=0.45\textwidth]{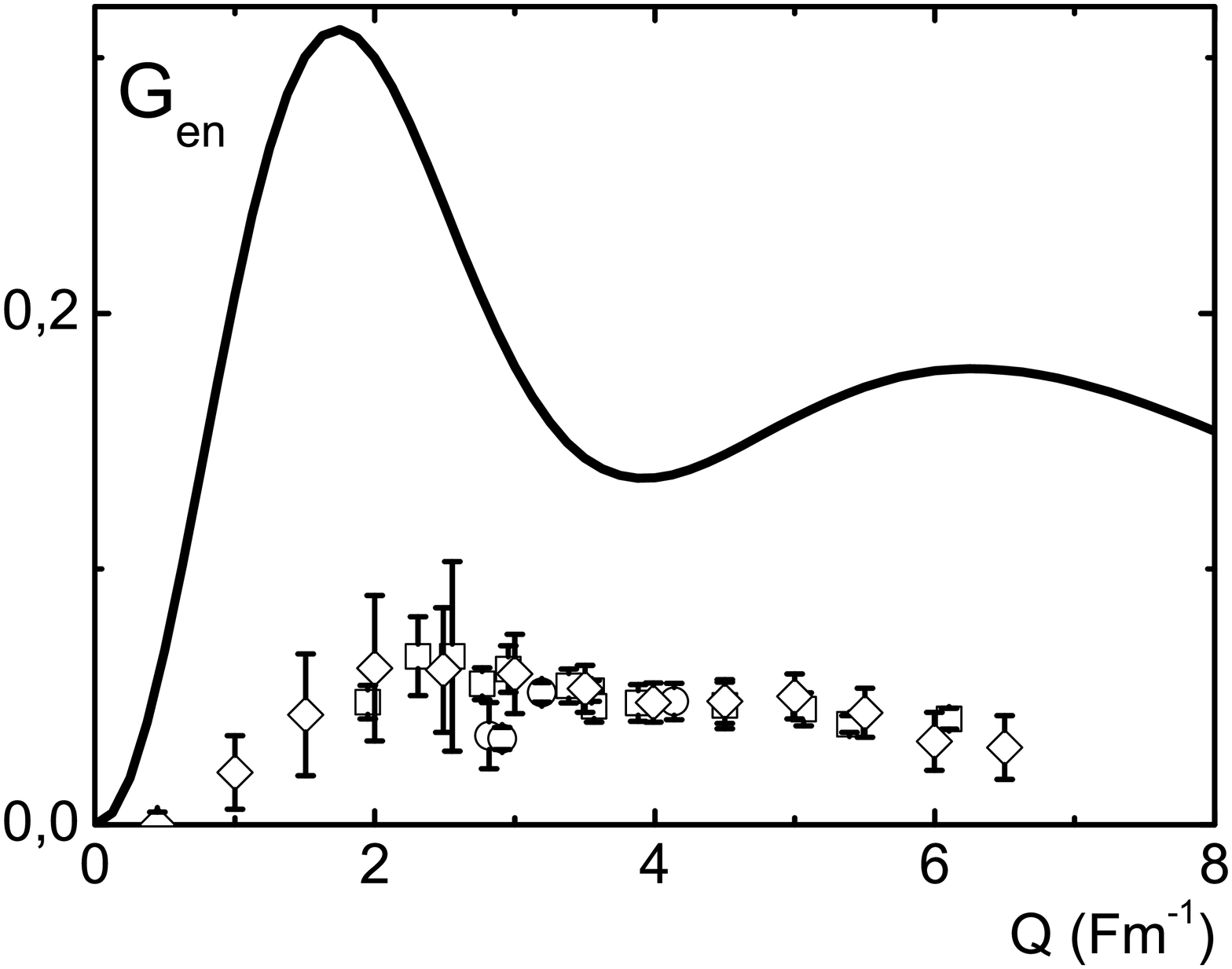}}
\centerline{\includegraphics[width=0.45\textwidth]{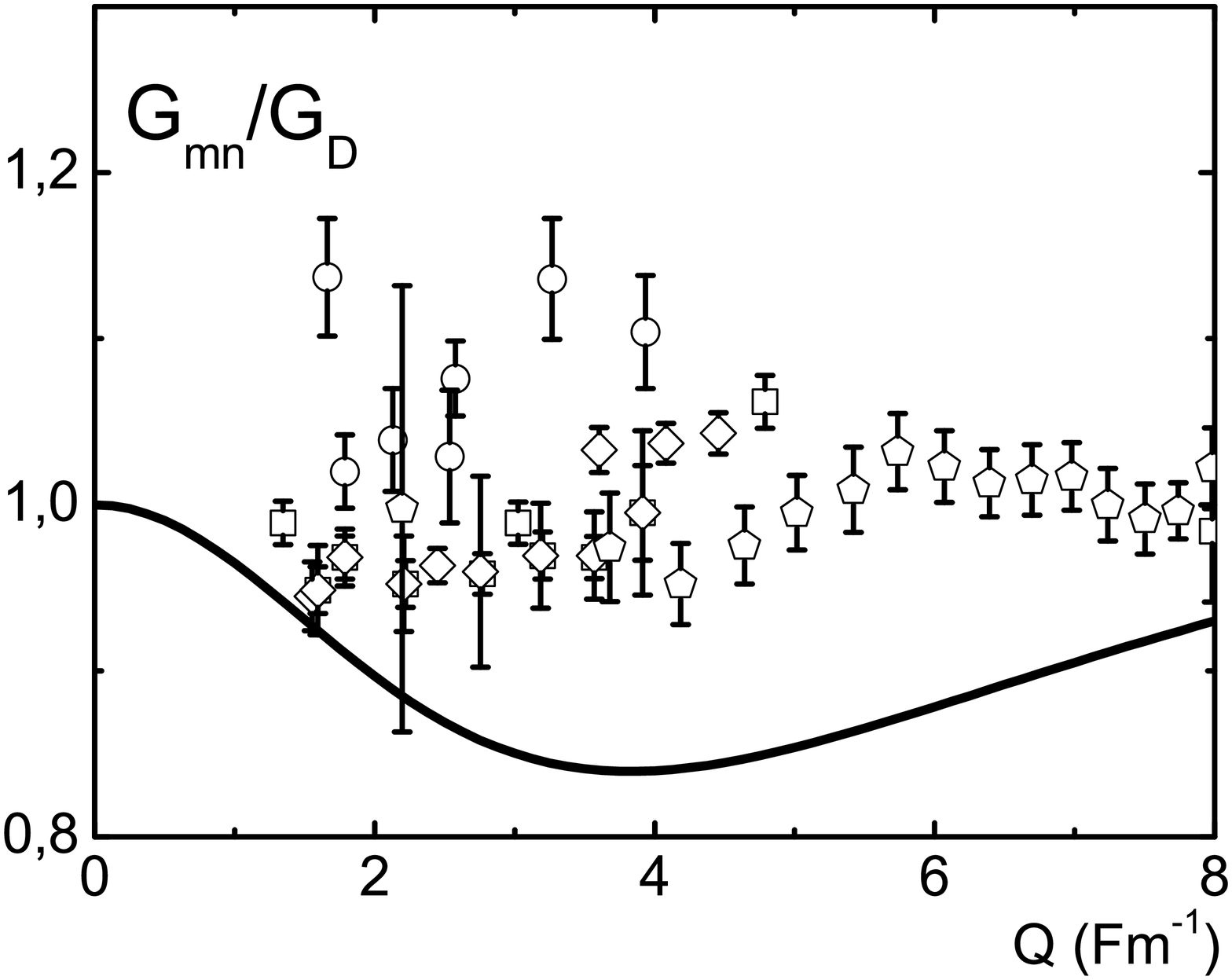}}
\caption{\label{fig:nFFs} Neuteron FFs as a function of Q. Data are compilation of \cite{MichaelKohl2008} 
of  various model dependent analyses. One must be aware that  these data were not extracted from direct experiment. }
\end{figure}
\begin{figure}[htb]
\centerline{\includegraphics[width=0.5\textwidth]{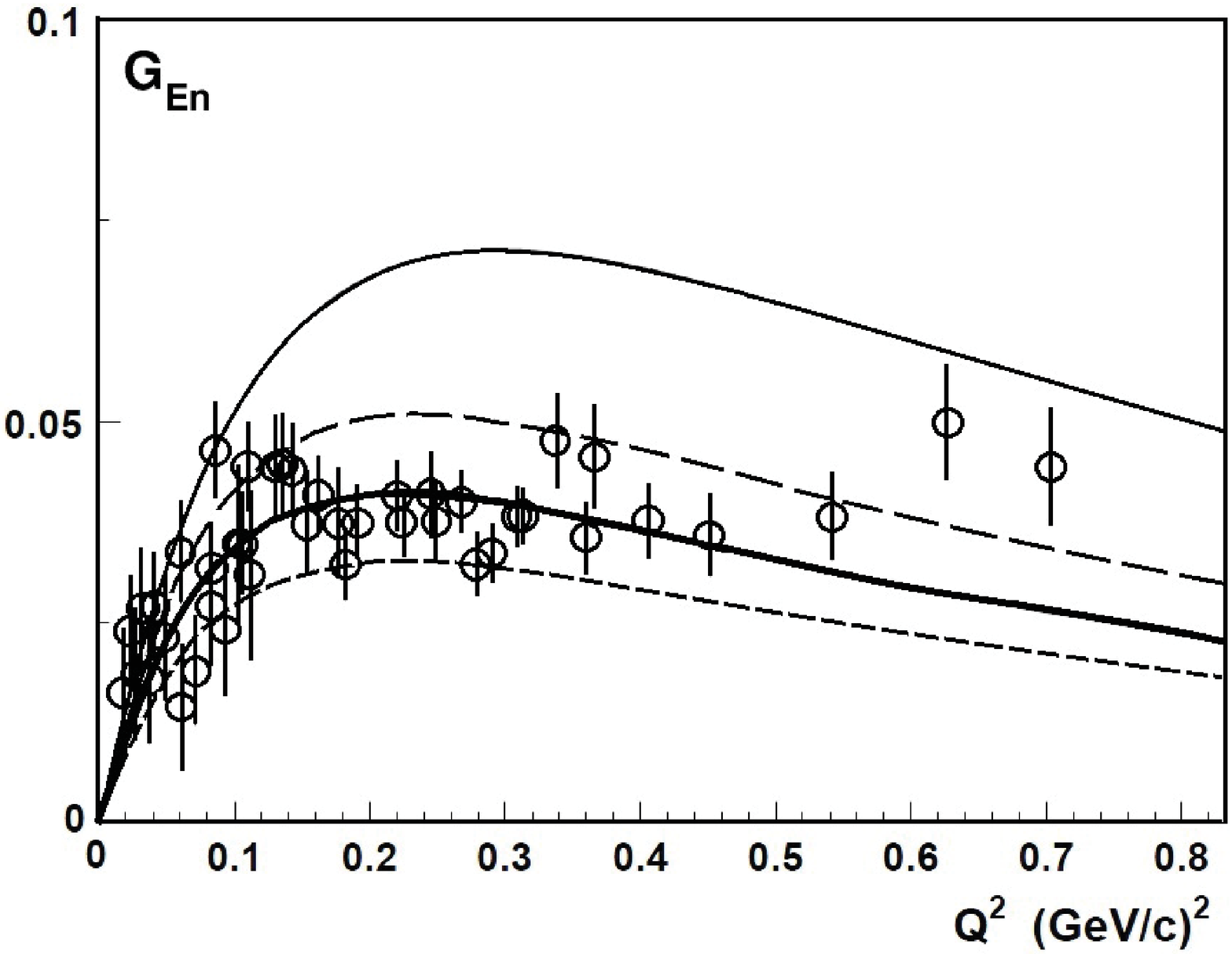}}
\caption{\label{fig:Plachkov} Data of Platchkov et al. \cite{Plachkov1} extracted with the
Paris potential. Lines are ¯fits to the same data extracted with Paris (thick solid), RSC
(short-dashed), Argonne (long-dashed), Nijmegen (solid) potentials. Figure is from \cite{Plachkov1}.}
\end{figure}

Our calculations show that the modest changes of the deuteron wave function and of
the nucleon form factor parameterizations may simulate
effects of the meson exchange currents for the eD elastic
scattering. The analysis of Plachkov et al \cite{Plachkov1} showed that the neutron electric FF extracted values are extremely model dependent (see Fig. 6). That analysis was made then data of polarization experiments were scare.
Unfortunately, there is no up-to-date analysis of the  neutron FFs data similar to one made in \cite{Plachkov1}.
This analysis is not easy to perform due to a number of the nucleus models (and exchange currents) used in the literature.
All the difficulties of the proton FFs extraction  are present here too, but have not been estimated so far.
Therefore data shown in all experimental papers must be considered as model results.
Experimental estimates of the neutron FFs error
bars are underestimated considerably, systematic errors
of the NN interaction uncertainties are not properly
accounted for. Our results for the nucleon FFs show that the extracted proton FFs are inside the experimental bars (fig.~4), but electric neutron FF (fig.~5) may be 2-3 times greater than results
extracted with some ``Arenh\"{o}vel's model'' \cite{Rioradan_2010}. In that case values of the magnetic neutron FF also change (fig.~5), but not dramatically.

We are going to calculate the deuteron electrodisintegration process to show that
this reaction may be described in the ``only NN channel'' relativistic model, and our preliminary estimates show the possibility.
\end{document}